\documentclass[conference]{IEEEtran}
\IEEEoverridecommandlockouts
\usepackage{cite}
\usepackage{amsmath,amssymb,amsfonts}
\usepackage{graphicx}
\usepackage{etoolbox}
\usepackage{textcomp}
\usepackage{xcolor}
\newcommand{\tabitem}{~~\llap{\textbullet}~~}
\usepackage{pgfplots}
\usepackage{tikz}
\usepackage{hhline}
\usepackage{color, soul}
\usepackage{multirow}
\usepackage[colorlinks=true,citecolor=black,linkcolor=black,urlcolor=blue]{hyperref}
\usepackage{tikz,graphics,color,float,epsf,caption,subcaption}
\usepackage{tikz}
\usepackage{todonotes}
\usetikzlibrary{shapes,arrows,calc,positioning}
\usepackage{amsmath,bm,times}
\tikzstyle{sensor}=[draw, fill=blue!20, text width=5em, 
    text centered, minimum height=2.5em]\tikzstyle{sensors}=[draw, fill=blue!20, text width=5em, 
    text centered, minimum height=2.5em]
\tikzstyle{ann} = [above, text width=5em]
\tikzstyle{naveqs} = [sensor, text width=3.0em, fill=blue!20, 
    minimum height=10em, rounded corners]
\tikzstyle{naveqss} = [sensors, text width=2.5em, fill=blue!20, 
    minimum height=2em, rounded corners]
\tikzstyle{sum} = [draw, circle, scale=0.7, node distance = 0.5cm]
\newcommand{\suma}{\small $+$}
\newcommand{\mula}{\small $\times$}
\newcommand{\expa}{\small $exp$}
\def\blockdist{2.0}
\def\edgedist{1.0}
\def\edgedista{1.5}
\usepackage{booktabs,threeparttable}

\makeatletter
\def\endthebibliography{%
  \def\@noitemerr{\@latex@warning{Empty `thebibliography' environment}}%
  \endlist
}
\makeatother

\def\BibTeX{{\rm B\kern-.05em{\sc i\kern-.025em b}\kern-.08em
    T\kern-.1667em\lower.7ex\hbox{E}\kern-.125emX}}
\begin{document}

\title{Multimodal Integration for Large-Vocabulary Audio-Visual Speech Recognition
\thanks{This project has received funding from the German Research Foundation DFG under grant number KO3434/4-2.}
}

\author{\IEEEauthorblockN{Wentao Yu, Steffen Zeiler, Dorothea Kolossa}
\IEEEauthorblockA{\textit{Institute of Communication Acoustics, Ruhr University Bochum, Germany} \\
\{wentao.yu, steffen.zeiler, dorothea.kolossa\}@rub.de
}

}

\maketitle

\begin{abstract}
For many small- and medium-vocabulary tasks, audio-visual speech recognition can significantly improve the recognition rates compared to audio-only systems. However, there is still an ongoing debate regarding the best combination strategy for multi-modal information, which should allow for the translation of these gains to large-vocabulary recognition. While an integration at the level of state-posterior probabilities, using dynamic stream weighting, is almost universally helpful for small-vocabulary systems, in large-vocabulary speech recognition, the recognition accuracy remains difficult to improve. In the following, we specifically consider the large-vocabulary task of the LRS2 database, and we investigate a broad range of integration strategies, comparing early integration and end-to-end learning with many versions of hybrid recognition and dynamic stream weighting. One aspect, which is shown to provide much benefit here, is the use of dynamic stream reliability indicators, which allow for hybrid architectures to strongly profit from the inclusion of visual information whenever the audio channel is distorted even slightly. 
\end{abstract}

\section{Introduction}
\label{sec:intro}
Large Vocabulary Continuous Speech Recognition
(LVCSR) remains difficult as a lipreading task, because many pairs of phonemes correspond to a single viseme, making many pairs of words almost indistinguishable to a vision-only system, as for example ``do'' and ``to''. Due to this intrinsic difficulty, an integration of lipreading into speech recognition becomes difficult in large- or open-vocabulary applications \cite{thangthai2018building}.
 Nonetheless, lip-reading gives a great benefit to human listening \cite{crosse2016eye}. In this work, we use an exemplary large-vocabulary dataset - the LRS2 corpus described in \cite{Afouras18c}, to test whether and how similar benefits are attainable for automatic systems. 

Many studies have shown that video information can dramatically improve small-vocabulary speech recognition performance, when the audio signals are recorded in a noisy environment. Often, stream weighting proved to be an effective method to combine audio and video information. As in \cite{meutzner2017improving}, separate models are then trained for each of the modalities, and possibly, for each of the feature streams per modality. Stream weighting is realized through a weighted combination of the DNN state posteriors of each modality
\begin{equation} \label{statefusion}
\textnormal{log }\widetilde{p}(s |\textbf{o}_t)=\sum_{i}^{}\lambda_t^i\cdot \textrm{log }{p}(s |\textbf{o}_t^{i}),
\end{equation}
where $\textrm{log }{p}(s |\textbf{\textrm{o}}_t^{i})$ is the log-posterior of state $s$ in stream $i$ and $\textnormal{log}\,\widetilde{p}(s |\textbf{\textrm{o}}_t)$ is its estimated combined log-posterior.
The stream weights $\lambda_t^i$ are typically predicted by appropriate reliability measures. In most of the state-based multi-modal integration studies, only two streams with few reliabilities are used. For example, in \cite{gurban2008dynamic} the weights are only estimated from an entropy estimate. In this work, we consider a broad range of possible reliability metrics and apply them to fuse the information of three models, one acoustic and two visual. 

Overall, we compare the performance of three different integration methods. The paper is organized as follows: Section \ref{e2evshybrid} discusses the differences between end-to-end and hybrid speech recognition models. Early integration and state-based integration are discussed  in  Section \ref{sysove}. Different reliability indicators are introduced in Section \ref{relia}.  Section \ref{experup}  explains  the experimental setup, while Section \ref{results}  shows the results. Finally, in Section \ref{clu}, we discuss the overall performance of all systems and give an outlook on future work.

\section{End-to-end vs. hybrid models}
\label{e2evshybrid}
In recent years, end-to-end acoustic speech recognition has quickly gained widespread popularity. In its original form, this model predicts character sequences directly from the audio signal. Different from the end-to-end model, in a hybrid uni-modal speech recognition system, an acoustic model is trained to calculate log-pseudo-posteriors $\textnormal{log}\ p(s |\textbf{o}_t)$. A decoder then uses these pseudo-posteriors to obtain the best word sequence by graph search through a language model \cite{bourlard2012connectionist}.

While the hybrid model has the disadvantage of higher complexity compared to end-to-end learning, many recent papers have still shown superior performance of hybrid ASR compared to end-to-end recognition, for example in \cite{luscher2019rwth}. 

For our application of multi-modal recognition, hybrid frameworks are advantageous for multi-modal fusion, because they allow for an integration at the level of the pseudo-posteriors, and for using reliability information, which has proven beneficial for multimodal integration in many studies, see e.g.~\cite{estellers2011dynamic,Abdelaziz2015,Zeiler2016}. End-to-end models, in contrast, typically use an attention mechanism rather than reliabilities for the multi-modal fusion, which is also the case in the baseline models that we consider \cite{sterpu_icmi18,Chung17}. Recently, \cite{sterpu2020teach} has extended the work of \cite{sterpu_icmi18} and improved the performance by using a loss function that explicitly considers facial action units. 

\section{System overview}
\label{sysove}
\subsection{System framework}
\label{systemframework}
\begin{figure} 
\centering
\noindent \begin{tikzpicture}
    \node (naveq) [naveqs] {\small Stream Integration Net};
    \path (naveq.65)+(-\blockdist,0) node (input1) {};
    \path (naveq.0)+(-\blockdist,0) node (input2) {};
    \path (naveq.-65)+(-\blockdist,0) node (input3) {};
    
    \path [draw, ->] (input1) -- node [above] {\small $\sigma _{t}^\mathrm{A}\ $} 
        (naveq.west |- input1) ;
    \path [draw, ->] (input2) -- node [above] {\small $\sigma _{t}^\mathrm{VA}\ \ $} 
        (naveq.west |- input2) ;
    \path [draw, ->] (input3) -- node [above] {\small $\sigma _{t}^\mathrm{VS}\ \ $} 
        (naveq.west |- input3) ;

    \draw [->] (naveq.65) -- node [above] {\small $\lambda_{t}^\mathrm{A}$} + (\edgedist,0) 
        node[sum, right] (suma1) {\small \mula};
    \draw [->] (naveq.0) -- node [above] {\small $\lambda_{t}^\mathrm{VA}$} + (\edgedist,0) 
        node[sum, right] (suma2) {\small \mula};
    \draw [->] (naveq.-65) -- node [above] {\small $\lambda_{t}^\mathrm{VS}$} + (\edgedist,0) 
        node[sum, right] (suma3) {\small \mula};
    \draw [->] ($(0,0.6cm)+(suma1)$)node[above]{\small $\textrm{log}\ p(s |\textbf{o}_t^\mathrm{A})$} -- (suma1);
    \draw [->] ($(0,0.6cm)+(suma2)$)node[above]{\small $\textrm{log}\ p(s |\textbf{o}_t^\mathrm{VA})$} -- (suma2);
    \draw [->] ($(0,0.6cm)+(suma3)$)node[above]{\small $\textrm{log}\ p(s |\textbf{o}_t^\mathrm{VS})$} -- (suma3);
    \node [sum, right = 0.5cm of suma2] (suma4) {\small \suma};
    \draw [->] (suma1) -| (suma4);
    \draw [->] (suma2) -- (suma4);
    \draw [->] (suma3) -| (suma4);
    \node [sum, right = 0.5cm of suma4] (exp) {\small \expa};
    
    \draw [->] (suma4) -- (exp);
    \draw [->] (exp) -- node [above] {\small $\widetilde{p}(s |\textbf{\textrm{o}}_t)$} + (\edgedista,0);

    
\end{tikzpicture}
\caption{Audio-visual fusion strategy for audio and two streams of different video models}
\label{fig:DNN}
\end{figure}
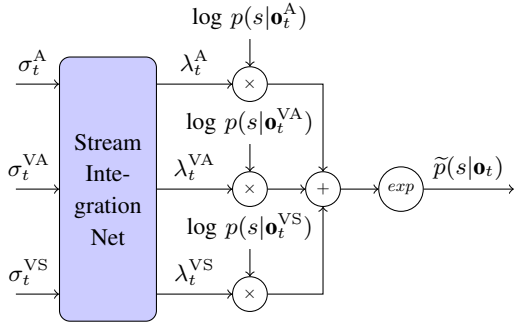

As shown in Fig. \ref{fig:DNN}, a set of reliability measures $\sigma _{t}^\mathrm{A}$, $\sigma _{t}^\mathrm{VA}$, $\sigma _{t}^\mathrm{VS}$ (described in more detail in Sec. \ref{relia}) is used as the input for the stream integration net, which uses these to obtain weights $\lambda_t^i$ for all streams over time. We then fuse the audio and video models through Eq.~\eqref{statefusion}. 

For training the stream integration net, we carry out forced alignment on the clean audio training set to obtain target state sequences $p^*(s | \textbf{\textrm{o}}_t)$, in which the reference state probability is set to one and any other state probabilities are set to zero. Two state-based loss functions are employed here, the cross-entropy (CE)
\begin{equation} \label{ce}
\textnormal{CE}= -\frac{1}{T}\sum_{t=1}^{T} \sum_{s=1}^{S}p^*(s | \textbf{\textrm{o}}_t)\cdot \textnormal{log }\widetilde{p}(s |\textbf{\textrm{o}}_t),
\end{equation}
 and the mean squared error (MSE)
\begin{equation} \label{mseeq}
\textnormal{MSE}= \frac{1}{T\cdot S}\sum_{t=1}^{T} \sum_{s=1}^{S}(p^*(s | \textbf{\textrm{o}}_t)- \widetilde{p}(s |\textbf{\textrm{o}}_t))^2.
\end{equation}
We also consider maximum mutual information (MMI) as a sequence-based criterion \cite{vesely2013sequence} via
\begin{equation}
\frac{\partial \mathit{F}_{MMI}}{\partial \textnormal{log }\widetilde{p}(s |\textbf{\textrm{o}}_t)} = \kappa (p^*(s | \textbf{\textrm{o}}_t)-\gamma^{DEN}_{t}(s)).
\end{equation}
Here, $\gamma^{DEN}_{t}(s)$ is the posterior probability of state $s$ at time $t$, computed over the denominator lattices that are obtained from the state pseudo-posteriors $\widetilde{p}(s |\textbf{\textrm{o}}_t)$. $\kappa$ is the acoustic scaling factor and set to $1.0$.
\subsection{Oracle weight baseline}
To get an estimate of the best achievable word error rate (WER), we use convex optimization via CVX \cite{cvx, gb08} to optimize the cross-entropy in  Eq.~\eqref{ce}, which yields a set of oracle stream weights.

\subsection{Early integration baseline}
Early integration fuses audio and video information directly at the input of the system, using stacked feature vectors via
\begin{equation} \label{eq:concat}
\textbf{\textrm{o}}_t=[(\textbf{\textrm{o}}_t^\mathrm{A})^T,(\textbf{\textrm{o}}_t^\mathrm{VS})^T,(\textbf{\textrm{o}}_t^\mathrm{VA})^T]^T
\end{equation}
where $\textbf{\textrm{o}}_t^\mathrm{A}$ are audio features, $\textbf{\textrm{o}}_t^\mathrm{VS}$ are shape-based video features, and $\textbf{\textrm{o}}_t^\mathrm{VA}$ are appearance-based video features, described in more detail in Sec. \ref{ssec:FE}, and $T$ denotes vector transposition. As the audio and video features have different frame rates, we use a digital differential analyzer, similar to Bresenham\textquotesingle s algorithm \cite{sproull1982using} to synchronize the video features before applying Eq.~\eqref{eq:concat}. 

\subsection{End-to-end baselines}
In addition to the early integration baseline, we compare the performance of our suggested hybrid audiovisual ASR to end-to-end models with attention mechanisms \cite{sterpu_icmi18,Chung17}, which offer an alternative approach to multimodal fusion.  As described in \cite{Chung17}, both audio and video encoders are LSTM networks. The decoder is an LSTM transducer \cite{bahdanau2014neural}, which uses the encoded audio and video sequences, either via a dual attention mechanism \cite{Chung17}, or, in \cite{sterpu_icmi18}, using a multi-head attention mechanism that is specifically optimized towards audio-visual integration.
\section{Reliability measures} \label{relia}
As shown in Tab. \ref{table:RMS}, the proposed reliability measures can be grouped into the ones that are model-based (\textbf{MB}) and the signal-based (\textbf{SB}) measures. For the model-based measures, the audio and video models are considered separately. The signal-based measures can also be subdivided into audio-based (\textbf{AB}) and video-based (\textbf{VB}) measures.

\begin{table}[H]
    \caption{Proposed reliability measures}
\label{table:RMS}
    \footnotesize
    \setlength\tabcolsep{2.0pt}

\begin{tabular}{|c|c|l|}
\hline
\multirow{3}{*}{Model-based (\textbf{MB})}                                                                                                                         & \multicolumn{2}{c|}{\multirow{2}{*}{Signal-based (\textbf{SB})}}                                                                                                                                                           \\
                                                                                                                                                     & \multicolumn{2}{c|}{}                                                                                                                                                                                        \\ \cline{2-3} 
                                                                                                                                                     & Audio-based                                                                                                                                 (\textbf{AB}) & \multicolumn{1}{c|}{Video-based (\textbf{VB})}                               \\ \hline
\multicolumn{1}{|l|}{\begin{tabular}[c]{@{}l@{}}\tabitem Entropy\\ \tabitem Dispersion \\ \tabitem Posterior difference\\ \tabitem Temporal divergence\\ \tabitem Entropy and \\ \ \ \ dispersion ratio\end{tabular}} & \multicolumn{1}{l|}{\begin{tabular}[c]{@{}l@{}}\tabitem MFCC\\ \tabitem $\Delta$MFCC\\ \tabitem SNR\\ \tabitem Signal and \\ \ \ \  noise energy\\ \tabitem Soft VAD\end{tabular}} & \begin{tabular}[c]{@{}l@{}}\tabitem IDCT\\ \tabitem Image distortion\end{tabular} \\ \hline
\end{tabular}
\end{table}

\subsection{Model-based reliability measures} 
The \textbf{\textit{entropy}} is a proxy for the model's uncertainty about the state $s$, given the current observation $\textbf{o}_t$. It is calculated for each stream $i$ via
\begin{equation} \label{entropy}
\textnormal{H}^i_t=-\sum_{s=1}^{S\,^i}p(s| \textbf{o}_t^i)\cdot \textnormal{log }p(s | \textbf{o}_t^i),
\end{equation}
with $S\,^i$ as the number of states in stream model $i$.

Similarly, the \emph{\textbf{dispersion}} is related to the decoder's discriminative power. It is computed by:
\begin{equation} \label{dispersion}
\textnormal{D}^i_t=\frac{2}{K(K-1)}\sum_{l=1}^{K}\sum_{m=l+1}^{K}\textnormal{log }\frac{\widehat{p}(l | \textbf{o}_t^i)}{\widehat{p}(m | \textbf{o}_t^i)},
\end{equation}
where the probabilities $p$ are sorted in descending order to obtain $\widehat{p}$. $K$ is set to 15.

The K-largest \emph{\textbf{posterior difference}}, defined via
\begin{equation} \label{diff}
\textnormal{Diff}^{\,i}_t=\frac{1}{K-1}\sum_{k=2}^{K}\textnormal{log }\frac{\widehat{p}(1 | \textbf{o}_t^i)}{\widehat{p}(k | \textbf{o}_t^i)},
\end{equation} 
is also considered, showing the average ratio between the largest posterior and the next $K-1$ values.

The \textit{\textbf{temporal divergence}} is computed as the Kullback-Leibler divergence between two posterior vectors $p(s | \textbf{o}_t^i)$ and $p(s | \textbf{o}_{t+\Delta t}^i)$, i.e.~
\begin{equation}
\textnormal{Div}_{\Delta t}^i(t)=\textnormal{D}_{KL}(p(s | \textbf{o}_t^i) ||p(s | \textbf{o}_{t+\Delta t}^i)).
\end{equation}
$\Delta t$ is set to $250\ $ms. As described in \cite{hermansky2013multistream}, the mean of the temporal divergence is also an interesting measure of reliability
and it is used here by averaging $\textnormal{Div}_{\Delta t}^i(t)$ over segments of $50\ $ms length.

The \textbf{\textit{entropy ratio}} is described in \cite{misra2003new}. The strongly related \textbf{\textit{dispersion ratio}} $\omega_{D,t}^i$ is estimated based on the average dispersion $\overline{\textnormal{D}}_t$
\begin{equation}
\omega_{D,t}^i= \frac{\widetilde{\textnormal{D}}_t^i}{\sum_{k=\mathrm{A,VA,VS}}^{}\widetilde{\textnormal{D}}_t^k},
\end{equation}
where 
\begin{equation}
\widetilde{\textnormal{D}}_t^i = \left\{\begin{matrix}
\frac{1}{10000} \ \ \textnormal{D}_t^i < \overline{\textnormal{D}}_t\\[2pt]
\ \ \ \textnormal{D}_t^i \  \ \ \ \textnormal{D}_t^i \geq \overline{\textnormal{D}}_t.
\end{matrix}\right.
\end{equation}
A, VA, and VS represent the audio, video appearance and video shape stream, respectively. $\textnormal{D}_t^i$ is obtained from Eq.~\eqref{dispersion}.
\vspace{-5pt}
\subsection{Signal-based reliability measures}
\label{signalrms}
The first 5 \textbf{\textit{MFCC}} coefficients and their temporal derivatives, \textbf{\textit{$\Delta$MFCC}}, are related to the audio quality.

The estimated Signal-to-Noise Ratio (\textbf{\textit{SNR}}) also represents the quality of the audio signal; it is computed in each frame
\begin{equation}
\mathrm{SNR\,_t}=10\ \textnormal{log}\left(\cfrac{\textnormal{S}_t}{\textnormal{N}_t}\right).
\end{equation}
The \textbf{\textit{signal energy}} $\textnormal{S}_t$ is estimated as the sum of squared amplitudes of the Hamming-windowed frame $t$. The \textbf{\textit{noise energy}} $\textnormal{N}_t$ is estimated by a variant of the minima-controlled recursive averaging algorithm (MCRA-2)\cite{rangachari2006noise}. 

The ratio between the energy of the speech band and the total energy of each frame is used as a soft voice-activity detection (\textbf{\textit{soft VAD}}) cue.

The first 5 Inverse Discrete Cosine Transform (\textbf{\textit{IDCT}}) coefficients of the mouth region represent low-level image properties.

The \textbf{\textit{image distortion}} measures comprise the lighting condition, the degree of blurring and the head pose, all computed as in \cite{schonherr2016environmentally}. The lighting condition represents the mean brightness of the image. The degree of blurring is estimated as the variance of the image after high-pass filtering. To obtain an indicator for head rotation and tilt, the cross-correlation between the original image and its horizontally mirrored version is computed.

\section{Experimental setup}
\label{experup}
\subsection{Dataset}
Our experiments are based on the LRS2 corpus. The training set contains 45,839 spoken sentences and 17,660 words, with a test set of 1,243 sentences and 1,698 words. To analyze the performance in different acoustic noise conditions, we have artificially created noisy versions of the LRS2 database
. The augmentation recipe from Kaldi's Voxceleb example 
is employed for this purpose, using the MUSAN corpus \cite{musan2015} as the noise dataset. It contains 3 different kinds of noise, i.e.~ambient noise, music and babble noise. Seven different SNRs are selected, from $-9$ dB to $9$ dB in steps of $3$ dB. Each audio signal is augmented with these three noise types and the SNR is randomly chosen from all SNRs. 
\subsection{Implementation details}
\label{sec:netsetup}
All hybrid recognition experiments are carried out using the Kaldi toolkit \cite{povey2011kaldi}, with the training set of the LRS2 corpus employed in the training of the acoustic and visual models. The initial HMM-GMM training follows the standard Kaldi AMI recipe; subsequent HMM-DNN training uses the nnet2 recipe. 
The output dimension of all three models is 3784. For performance reasons, the audio model alignments are also used for the HMM-DNN training of both video models. The integration model has 5 hidden layers, with 43, 25, 17, 10 and 3 units, respectively, each using ReLU activation functions. The output is the predicted weight of each stream, $\lambda_t^i$. A sigmoid function is used to limit the weights to values between 0 and 1. Finally, we normalize the multi-modal posterior probabilities to one at each time $t$. To avoid overfitting, early stopping is used if the training loss does not improve for 1200 iterations. The stream integration model is trained on the development set and performance is tested on the evaluation set.
\subsection{Feature extraction}
\label{ssec:FE}
The audio model uses 13-dimensional MFCCs as features. MFCCs are extracted with
$25$ ms frame size and $10$ ms frame shift. The video frame is 40 ms long without overlap. The mouth region is detected via OpenFace \cite{amos2016openface}. The video appearance model (VA) uses 43-dimensional IDCT coefficients of the mouth region in the grayscale image as features. The video shape model (VS) uses the 34-dimensional non-rigid shape parameters \cite{amos2016openface} as features. 
 
\section{results}
\label{results}
Here, we provide the performance comparisons between our proposed hybrid model and the end-to-end AVSR models, which are described in \cite{Chung17} and \cite{sterpu_icmi18}.

\begin{figure}[htb]
    \includegraphics[width=8cm,height=12cm,keepaspectratio]{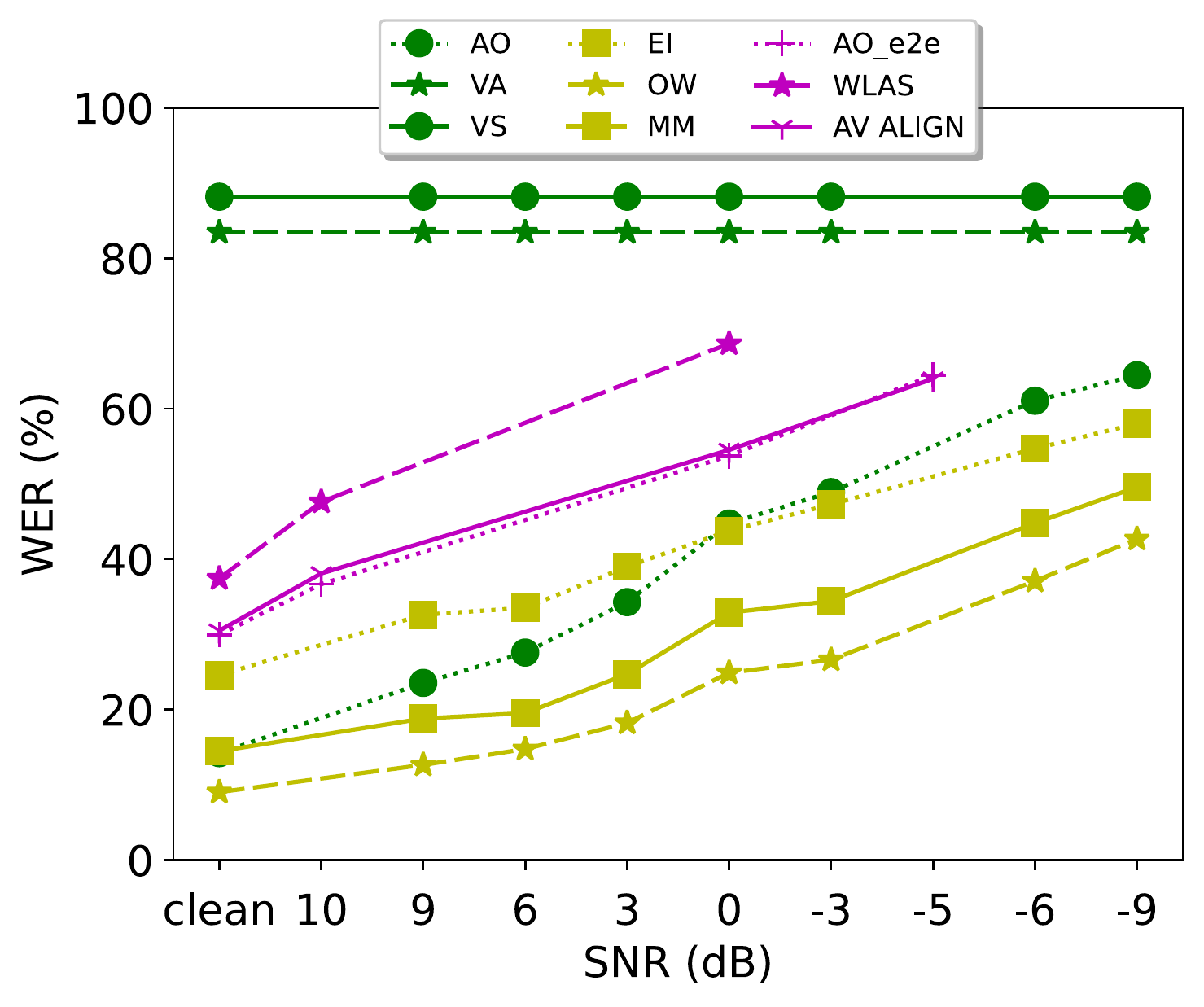}
    \caption{Word error rate on LRS2 corpus}
\label{fig:baseline}
\end{figure}

Fig.~\ref{fig:baseline} shows the performance of all considered baseline hybrid models (more details in Tab.~\ref{table:results}). The audio-only model (\textbf{AO}) has a much better performance than the video-shape (\textbf{VS}) and video-appearance (\textbf{VA}) models alone. Early integration (\textbf{EI}) can already improve the WER at lower SNR conditions ($\leqslant 0$ dB), but there is no improvement, if we compare the average WER over all SNRs. As the oracle weight model (\textbf{OW}) shows, there is much room for improvement through optimal stream integration.

Fig.~\ref{fig:baseline} also depicts the word error rate results of different end-to-end models, which can be found in the appendix of \cite{sterpu_icmi18}. The Watch Listen Attend and Spell model \cite{Chung17} (\textbf{WLAS}) and the proposed fusion strategy in \cite{sterpu_icmi18} (\textbf{AV Align}) can not improve the WER on the LRS2 dataset. We also find that the hybrid audio-only model (\textbf{AO}) offers a better performance than the end-to-end audio-only model (\textbf{AO{\_}e2e}) from \cite{sterpu_icmi18}.

\begin{threeparttable}
    \caption{WER (\%) on LRS2.}
\label{table:results}
    \footnotesize
\renewrobustcmd{\bfseries}{\fontseries{b}\selectfont}
\renewrobustcmd{\boldmath}{}
\newrobustcmd{\B}{\bfseries}
    \setlength\tabcolsep{2.5pt}
    \renewcommand{\arraystretch}{1.2}
\begin{tabular}{cccccccccc}
\toprule
SNR  & -9    & -6    & -3    & 0     & 3     & 6     & 9     & clean & avg.\\
\hline
AO    & 64.45 & 61.05 & 48.90 & 44.73 & 34.28 & 27.57 & 23.54 & 14.20 & 39.84\\
VA  & 83.44 & 83.44 & 83.44 & 83.44 & 83.44 & 83.44 & 83.44 & 83.44 & 83.44\\
VS & 88.18 & 88.18 & 88.18 & 88.18 & 88.18 & 88.18 & 88.18 & 88.18 & 88.19\\
\hline
EI & 58.01 & 54.69 & 47.27 & 43.77 &38.99 &33.53 &32.60 &24.58 &41.68\\
\hline
OW   & 42.67 & 37.08 & 26.61 & 24.88 & 18.22 & 14.74 & 12.64 & 9.02 & 23.23\\
\hline
MSE &50.66 &47.10 &35.81 &33.56 &25.20 &19.93 &\footnotesize \B 18.29 &13.60 &30.52\\
CE &50.90 &48.40 &36.17 &33.94 &25.47 &19.86 &18.36 &\footnotesize  \B 13.45 &30.82\\
MMI &51.69 &48.74 &36.53 &33.75 &25.85 &20.11 &18.87 &13.60 &31.14\\
MM &\footnotesize \B 49.58 &\footnotesize \B 44.78 &\footnotesize \B 34.41 &\footnotesize \B 32.88 &\footnotesize \B 24.70 &\footnotesize \B 19.53 &18.80 &14.46 &29.89\\

\hline
\\
\end{tabular}
\end{threeparttable}

Tab.~\ref{table:results} summarizes all results of the Kaldi experiments: the first 3 rows show the performance of all single-modality models. The performance metrics of the video appearance and shape models are far from satisfying. We have also employed the pre-trained spatio-temporal visual front-end from \cite{stafylakis2017combining} to extract high-level visual features, without seeing improvements.  We hypothesize that the unsatisfying video model performance is due to an insufficient amount of training data.

 The 4th and 5th row show the results of the early-integration baseline, and of the oracle weighting that gives an upper bound of achievable performance for the considered hybrid architecture. The final 4 rows show the WERs for our proposed experiments, using all reliability indicators. Comparing the different loss functions for training the stream integration net, cf.~Sec.~\ref{systemframework}, the mean squared error (\textbf{MSE}) has the best performance at lower SNR conditions ($\leqslant 3$ dB), which can be carried over into the sequence-based optimization by adding an MSE-based fine-tuning after pre-training with the original \textbf{MMI} loss (\textbf{MM}). Comparing the best performance between our proposed hybrid audio-visual model (\textbf{MM}) and the end-to-end audio-visual model (\textbf{AV ALIGN}) in Fig.~\ref{fig:baseline}, we find that the hybrid model offers clear performance benefits, and that, in contrast to the end-to-end integration mechanism, it is indeed able to profit strongly from the inclusion of the visual modality under all noisy conditions.

To compare the value of the different types of reliability indicators for multi-modal integration, we have repeated this experiment with different reliability sets, using the best combination of loss functions, MMI training with MSE fine-tuning (\textbf{MM}).
Tab.~\ref{table:resultsreabi} gives the average WER over all SNR conditions. Here, we find that audio-based reliabilities have a slightly higher benefit, but using all reliabilities simultaneously achieves the best performance overall, corresponding to a relative word-error-rate reduction of 24.97\% over the best audio-only model.
\vspace{-5pt}
\begin{threeparttable}[htb]
    \caption{WER (\%) of different reliability sets,  abbreviations as introduced in Tab. \ref{table:RMS}}
\label{table:resultsreabi}
    \small
\renewrobustcmd{\bfseries}{\fontseries{b}\selectfont}
\renewrobustcmd{\boldmath}{}
\newrobustcmd{\B}{\bfseries}
    \setlength\tabcolsep{9.3pt} 
\begin{tabular}{cccccc}
\toprule
              & MB    & SB     & AB    & VB   & All \\
MM   & 31.40 & 31.33 & 31.03 & 31.76 &\B 29.89 \\
\hline
\\
\end{tabular}
\end{threeparttable}
\vspace{-5pt}

\section{conclusion}
\label{clu}

Improving the performance of large-vocabulary speech recognition through the inclusion of video data has remained challenging despite much progress in deep learning models for speech recognition and image processing. In this paper, we address this issue by learning an explicit stream integration network for audio-visual speech recognition. This network utilizes stream reliability indicators to optimize stream fusion time-frame by time-frame, ultimately providing a discriminatively optimized fusion of state-posteriors for hybrid speech recognition. We have compared the performance of this learned integration model to that of early integration as well as to a baseline end-to-end model. All experiments show that the proposed model dramatically outperforms both of these baseline systems and that it is capable of providing clear improvements in accuracy compared to audio-only recognition, as hoped.

However, experiments based on oracle knowledge for stream fusion also point at the possibility of significant further gains. Achieving similar improvements without oracle information will be the natural next goal of our work, where we will focus on both the topology and loss function of the fusion network as well as on the integration of deeper, pre-trained image recognition models.

\bibliographystyle{IEEEbib}

\bibliography{refs}


\end{document}